\documentclass[aps,prb,twocolumn,showpacs,superscriptaddress]{revtex4}
\usepackage{graphics}
\usepackage{longtable}
\usepackage[epsfig]{graphicx}
\usepackage{epsfig}
\begin{document}

\title{CH$_x$ adsorption ($x$=1-4) and thermodynamic stability on the CeO$_2$(111) surface: \\A first-principles investigation}

\author{Marco \surname{Fronzi}}
\email[Corresponding author.\\ E-mail: ]{marco@gmail.com}
\affiliation{Department of Mechanical Science and Bioengineering, Graduate School of Engineering Science, Osaka University, Japan}

\author{Simone \surname{Piccinin}}
\affiliation{CNR-IOM Democritos, c/o SISSA, via Bonomea 265, I-34136 Trieste, Italy}

\author{Bernard \surname{Delley}}
\affiliation{Paul-Scherrer-Institut, CH-5232 Villigen PSI, Switzerland}

\author{Enrico \surname{Traversa}}
\affiliation{Physical Sciences and Engineering Division, King Abdullah University of Science and Technology , Kingdom of Saudi Arabia}

\author{Catherine \surname{Stampfl}}
\affiliation{School of Physics, The University of Sydney, Sydney, New South Wales 2006, Australia}

\date{\today}

\begin{abstract}

We present an \emph{ab initio} investigation of the interaction between methane, its dehydrogenated forms and the cerium oxide surface. In particular, the stoichiometric CeO$_2$(111) surface and the one with oxygen vacancies are considered. We study the geometries, energetics and electronic structures of various configurations of these molecules adsorbed on the surface in vacuum, and we extend the analysis to realistic environmental conditions. A phase diagram of the adsorbate-surface system is constructed and relevant transition phases are analyzed in detail, showing the conditions where partial oxidation of methane can occur.

\end{abstract}





\maketitle

\section{Introduction}

The fuel cell technology developed in the past years has become an area of great interest, due to the high potential for the production of clean energy.\cite{long2013, kendrick2013} The research in this field aims to produce more efficient materials to be used as electrodes with a high and selective catalytic activity, in order to improve the efficiency of energy production and also to reduce the operating temperature of these devices.   
Therefore, recently many studies have been performed in order to analyze molecule-surface interactions and surface catalytic activity of several metal oxide compounds from both an experimental and a theoretical point of view.\cite{Moussounda2007, monnerat2001}

Cerium oxide has been considered a good candidate as an electrolyte material and an electrode support due to its desirable characteristic of storing oxygen, as well as its surface catalytic activity.\cite{saraf2005, kim2004, trovarelli1996, trovarelli2002, farrauto2012} Its surfaces show activity with respect to several chemical reactions, in particular for the C-H bond cracking of the methane molecule, proving that CeO$_{2-x}$ acts as active center for methane oxidation. \cite{haneda1998} 
This makes it a good candidate to be used as a fuel cell anode material.\cite{laosiripojana2004, knapp2008}
On the other hand, methane is the main component of natural gas and there is a keen interest to understand the mechanisms of methane oxidation, used for energy production.\cite{mcintosh2004,murray1999}
Direct oxidation of methane on both metal and metal-oxide surfaces has been observed, showing a very high thermodynamic efficiency limit for complete oxidation, that can reach 92$\%$.\cite{knapp2008,odier2007, bharadwaj1995}
Indeed, experiments proved that methane shows chemical activity if exposed to an atmosphere containing oxygen and that a humid environment play an important role against methane oxidation. \cite{cullis1984, ribeiro1994, kikuchi2002} 
 From this prospective, a systematic study of CH$_x$ adsorption on cerium oxide surfaces under realistic catalytic conditions becomes crucial to improve the efficiency of important processes in chemical industries.

 Several theoretical computational studies have been carried out to study methane dissociation on CeO$_2$ under vacuum conditions.\cite{knapp2008, mayernick2008} In this paper we want to examine this system when in contact with two gas reservoirs, oxygen and water vapor, by means of \emph{ab initio} atomistic thermodynamics, in order to reproduce realistic environmental conditions.\cite{stampfl2005, reuter2002, reuter2003, stampfl2002} 
   
Using a computational Density Functional Theory (DFT) approach, in this work we provide a systematic analysis of CH$_x$ adsorption on the stoichiometric, as well as the reduced, CeO$_2$(111) surface. We analyze the CH$_x$ fragment adsorption energies in vacuum, and describe the adsorbate-surface bonds obtained by charge distribution in the adsorption process and the electronic structures of the adsorbate-surface system.
The calculated adsorption and surface energies enable us to describe the trend of the Gibbs free energy under realistic catalytic conditions as a function of the chemical potential of the oxygen and water.
 The analysis results in the construction of a surface phase diagram that shows the relative thermodynamic stability of each adsorbate-surface system; also, relevant transition phases can be analyzed in detail.

\section{Details of Methodology}

\subsection{DFT calculations: basis set and convergence}

All calculations presented in this work are performed using the generalized gradient approximation for the exchange and correlation potential due to Perdew, Burke and Erzerhof (PBE)
 as implemented in the all-electron DMol$^{3}$ code.\cite{perdew1996,delley1990, delley2000} The DMol$^{3}$ method employs fast converging three-dimensional numerical integrations to calculate the matrix elements occurring in the Ritz variational method. The wave functions are expanded in terms of a double-numerical quality localized basis set with a real-space cutoff of 11 Bohr. Polarization functions and scalar-relativistic corrections are incorporated explicitly. 
We use (2$\times$2) surface unit cells, and a vacuum  region of $\approx$ $30$ \rm{\AA}, which ensures negligible interaction between periodic replicas, and an $8\times8\times1$ {\bf k}-point mesh, yielding 10 {\bf k}-points in the irreducible part of the Brillouin Zone. 
Further details about the use of DFT on systems containing reduced CeO$_2$ can be found elsewhere.\cite{fronzi2009, fronzi2009-bis}

\section{Geometries and Energetics}

In this section, we analyze energetics and geometries of the interaction between the CH$_x$ ($x$=1-4) and the CeO$_2$(111) surface with and without surface oxygen vacancies, when the CH$_x$ coverage is 0.25 ML under vacuum conditions. We start by considering the interaction of CH$_4$ with the CeO$_2$(111) surface. 
Various configurations are considered for adsorption, as reported in Tab.\ref{tab1}, where the adsorption energies are calculated as follows:

\begin{equation}
	 E{_{\rm ads}} = E{_{\rm sys}}-E{_{\rm surf}}-E{_{\rm mol}},
	\label{ads}
\end{equation}


where $E{_{\rm surf}}$ is the surface energy, $E{_{\rm mol}}$ the isolated molecule energy, and $E{_{\rm sys}}$ is the surface-adsorbate system.

\begin{table}[!tbp]
\caption{\label{tab1} Adsorption energies of CH$_4$ on different adsorption sites of the CeO$_2$(111) surface. Namely, on top of the Ce atom, O atom, surface oxygen vacancy (V$_{\rm {O}}$) and the oxygen atom located in the second layer (O$_{\rm {2nd}}$).  }
\begin{ruledtabular}   
\begin{tabular}{c|ccccccc}

Config.    &Adsorption energy  (eV)              \\

\hline

Ce            &     $-$0.07                                      \\
 
O           &   $-$0.05                               \\

O$_{\rm {2nd}}$    &  $-$0.06          \\

 V$_{\rm {O}}$   & 0.00          \\

\end{tabular}
\end{ruledtabular}
\end{table}

\begin{figure}[!tbp]
\scalebox{0.35}{\includegraphics{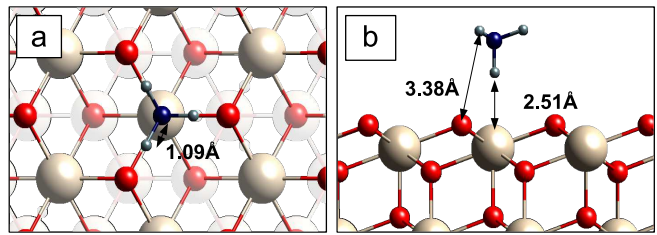}}
\caption{\label{fig:1} (Color online) Top (a) and side (b) view of adsorption sites considered for methane on the CeO$_2$(111). (a) and (b) show methane in the Ce configuration.  Large pale (gray) and small dark (red) spheres indicate Ce and O atoms, respectively. The small dark (blue) and the very small (gray) spheres represent C and H atoms respectively. Selected inter-atomic distances are indicated.}
\end{figure}

For the considered configurations, the methane molecule shows a weak interaction with the CeO$_2$(111) surface. In the most favorable case, it adsorbs at the cerium site, with an adsorption energy of $-$0.07 eV. In this configuration, shown in Fig. \,\ref{fig:1}, the methane molecule adsorbs on the surface with a hydrogen atom pointing to the surface cerium atom. The distance Ce-O in this case is 2.51 {\AA}. In this case, the energy value of the adsorbate-surface bond is very small, showing an extremely weak interaction. This result is in line with other results in the literature, using both DFT and DFT+U approaches.\cite{mayernick2008}  
When an oxygen vacancy is created on the cerium oxide surface, interestingly, the interaction become repulsive. The morphology of the surface becomes therefore crucial for the first step in the methane oxidation reaction.
To better analyze the nature of the interaction between CH$_4$ and the surface cerium atom, we studied the charge density difference ($Q_d$) induced by the adsorption of water on CeO$_2$(111), defined as the difference between the charge density of the adsorbate system ($Q_{s+a}$) and the sum of the isolated molecule ($Q_{a}$) and the slab ($Q_{s}$), calculated as follows:

\begin{equation}
 Q_d = Q_{s+a} - (Q_s + Q_a).
\label{charge} 
\end{equation}

In this case, an extremely small charge redistribution is observed (therefore not reported here). The adsorption energy in the proximity of an oxygen vacancy shows a slightly repulsive interaction between methane molecule and surface.

\begin{figure}[!tbp]
\scalebox{0.42}{\includegraphics{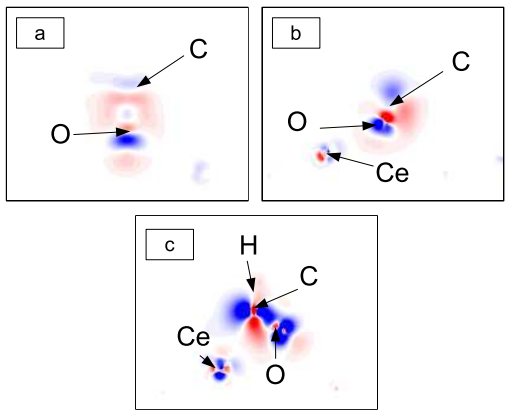}}
\caption{\label{charge_chx.eps} (Color online) Induced charge density due to the adsorption of (a) CH$_3$, (b) CH$_2$ and (c) CH on the CeO$_2$(111) surface. The red areas indicate charge accumulation, while the blue areas indicate charge depletion.}
\end{figure}

In order to obtain a first insight in the analysis of the CH$_4$ dissociation on the CeO$_2$(111) surface, we examine the adsorption of CH$_x$ fragments.
The CH$_3$ fragment adsorbs on the oxygen atom site with an adsorption energy of $-$2.09 eV, and the distance between carbon atom and surface oxygen is 1.40 {\AA} (see Tab.\ref{tab2}). The adsorption on the cerium atom and over the oxygen in the second layer are highly unfavored and the energy indicates a repulsive interaction.
When an oxygen vacancy is created on the surface, the adsorption energy decreases to $-$1.80 eV, indicating a less favorable interaction between molecule and reduced surface. 
The induced charge density of CH$_3$ adsorption on the CeO$_2$(111) indicates a charge redistribution, as shown in Fig.\ref{charge_chx.eps}, which shows an evident charge transfer from the surface oxygen toward the carbon. The charge accumulation is not strongly localized, likely due to the presence of an unpaired electron of the carbon atom. However, the shared charge can be an indication of a covalent bond between the carbon atom of the fragment and the surface oxygen.

\begin{table}[!tbp]
\caption{\label{tab2} adsorptionenergies (in eV) of CH$_3$ on the CeO$_2$(111) surface in different configurations. Namely, on top of the O atom and the surface oxygen vacancy (V$_{\rm {O}}$).}
\begin{ruledtabular}   
\begin{tabular}{c|cccccc}

Config.    &adsorption energy                \\

\hline


O       &   $-$2.09                                      \\


 V$_{\rm {O}}$  & $-$1.80                               \\

\end{tabular}
\end{ruledtabular}

\end{table}



\begin{figure}[!tbp]
\scalebox{0.35}{\includegraphics{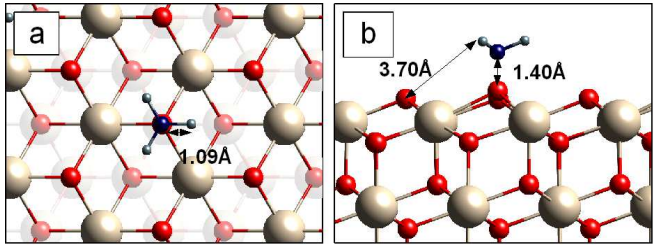}}
\caption{\label{fig:2} (Color online) Top (a) and side (b) view of adsorption sites considered for CH$_3$ fragment on  CeO$_2$(111). (a) and (b) show CH$_3$ in the O configuration.   The small dark (blue) and the very small (gray) spheres represent C and H atoms respectively.  Selected inter-atomic distances are indicated.}
\end{figure}


The CH$_2$ fragment interacts strongly with the surface, having a strong adsorption energy  on both the Ce and O sites ($-$3.26 eV and $-$3.31 eV, respectively). Even when it adsorbs on an oxygen vacancy, the adsorption energy differs from the most stable configuration only by 9.3$\%$ of the energy bond (see Tab.\ref{tab3}). 
The analysis of the induced charge density indicates an evident charge transfer from the surface oxygen to the carbon atom. The shared electron charge suggests the formation of a covalent bond.

\begin{table}[!tbp]
\caption{\label{tab3} Adsorption energies (in eV) of CH$_2$ on the CeO$_2$(111) surface in different configurations. Namely, on top of the Ce atom, O atom and surface oxygen vacancy (V$_{\rm {O}}$).}
\begin{ruledtabular}   
\begin{tabular}{c|ccccc}

Config.    &adsorption energy  (eV)              \\

\hline

Ce        &  $-$3.26                                        \\

O        &   $-$3.31                                      \\

 V$_{\rm {O}}$      &  $-$3.00                                     \\

\end{tabular}
\end{ruledtabular}

\end{table}


\begin{figure}[!tbp]
\scalebox{0.35}{\includegraphics{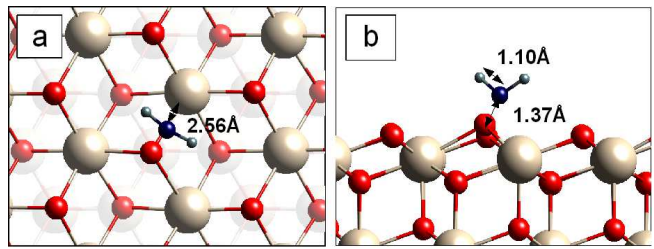}}
\caption{\label{fig:3} (Color online) Top (a) and side (b) view of adsorption sites considered for CH$_2$ on the CeO$_2$(111). (a) and (b) show CH$_2$ in the O configuration.  Large pale (gray) and small dark (red) spheres indicate Ce and O atoms, respectively.  The small dark (blue) and the very small (gray) spheres represent C and H atoms respectively.  Selected inter-atomic distances are indicated.}
\end{figure}


The last fragment analyzed in this section is CH. In this case the adsorption energy is very high when it adsorbs on both the Ce and O sites ($-$4.07 eV and $-$4.14 eV, respectively) and when an oxygen vacancy is created on the surface, the interaction energy decreases to $-$2.96 eV, indicating a strong interaction between the oxygen vacancy and the CH fragment. The induced charge density for the adsorption on the O site of CeO$_2$(111) indicates an evident charge transfer. Figure \ref{charge_chx.eps} shows a charge depletion around the surface oxygen and around the carbon atom in the plane cutting the aligned C,O Ce and H atoms. To obtain information about the bond formation, more detailed three dimensional charge distribution around the carbon atom is necessary. However, the evident charge redistribution around the cerium atom suggests a possible interaction between the carbon atom and the lower lying cerium atom.

\begin{table}[!tbp]
\caption{\label{tab4} adsorptionenergies (in eV) of CH on the CeO$_2$(111) surface in different configurations. Namely, on top of the Ce atom, O atom and surface oxygen vacancy (V$_{\rm {O}}$). }
\begin{ruledtabular}   
\begin{tabular}{c|ccccc}

Config.    &adsorption energy  (eV)              \\

\hline

Ce        &  $-$4.07                                        \\

O        &   $-$4.14                                      \\

 V$_{\rm {O}}$       &  $-$2.96                                     \\

\end{tabular}
\end{ruledtabular}

\end{table}



\begin{figure}[!tbp]
\scalebox{0.35}{\includegraphics{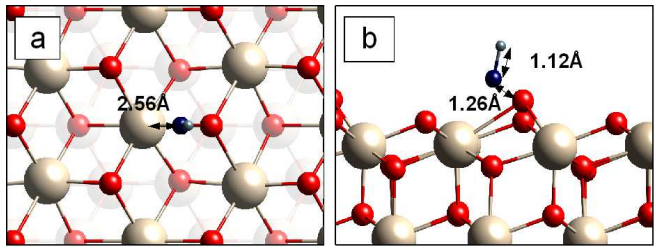}}
\caption{\label{fig:4} (Color online) Top (a) and side (b) view of adsorption sites considered for CH fragment on CeO$_2$(111). (a) and (b) show CH in the O configuration. Large pale (gray) and small dark (red) spheres indicate Ce and O atoms, respectively. The small dark (blue) and the very small (gray) spheres represent C and H atoms respectively.  Selected inter-atomic distances are indicated.}
\end{figure}


\section{Energetics and thermodynamics}

In this section we analyze the stability of the surfaces under the realistic environmental conditions of a humid atmosphere (H$_2$O) containing pure oxygen gas. Each adsorbate-surface system, previously considered in a vacuum space, is now analyzed when in thermodynamic equilibrium with two independent reservoirs of oxygen gas and water in its vapor phase.  We take into consideration the situation where the CeO$_2$(111) surface has one of the CH$_x$ fragment adsorbed, and where the oxygen and water chemical potentials can hypothetically vary independently to each other within two extreme limits of poor and rich concentration. This approach allows us to calculate the energy of each surface-molecule system under every possible value of oxygen/water partial pressure. 
The surface free energy for each of the surface-molecule systems is calculated as follows:

\begin{equation}
\gamma(\{p{_i}\}, T)  = \frac{1}{2A}  \left[ G - \sum N{_i} \mu{_i}(p_{i},T) \right],
\label{gamma}
\end{equation}

where $G$ is the Gibbs free energy of the solid with surface area
$A$ (the factor $\frac{1}{2}$ is due to the presence of two identical surfaces,
one on each side of the slab); $\mu$$(p{_i},T)$ is the chemical potential of the species $i$ that depends on pressure and temperature of the system, and
$N_i$ is number of atoms of component $i$ of the system.

The chemical potentials, $\mu(p,T)$, of oxygen and water vapor
depend on temperature and pressure according to:

\begin{equation}
\mu{_ {\rm O}} (p,T) = \frac {1}{2} \left[ E_{\rm O_{2}} + \widetilde{ \mu}_{\rm O_{2}} (p^{0},T) + k_{B}Tln \left( \frac{p_{\rm O_{2}}}{p^{0}} \right)  \right]
\label{chempot1}
\end{equation}

\begin{equation}
\mu_{\rm  H_{2}O} (p,T) =  E_{\rm H_{2}O} + \widetilde{\mu}_{\rm H_{2}O} (p^{0},T) + k_{B}Tln \left( \frac{p_{\rm H_{2}O}}{p^{0}} \right)  
\label{chempot2}
\end{equation}
 
\begin{figure}[!tbp]
\scalebox{0.28}{\includegraphics{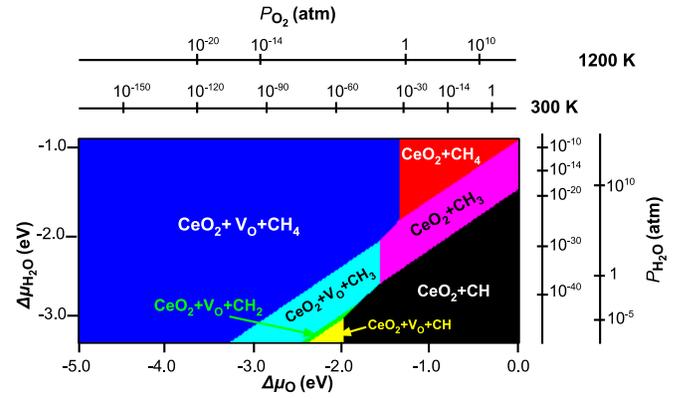}}
\caption{\label{fig:stability} Surface phase diagram of stable structures of CeO$_2$(111)+CH$_x$ in equilibrium with
a humid environment, as a function of $\Delta \mu_{\rm O}$ and $\Delta \mu_{\rm H_{2}O}$ in the gas phase. The additional axes
show the corresponding pressure scales at $T$=300 and 1200 K.}
\end{figure}

In Fig.\ref{fig:stability} these quantities are re-defined in order to have 0 eV as oxygen and water rich-limit conditions, and the plot is constructed as a function of $\Delta \mu_{\rm O_{2}}$ and $\Delta \mu_{\rm H_{2}O}$.  
For this specific system, the poor-limit conditions for the chemical potentials are $ \Delta \mu_{\rm O_{2}} > -5.01 $ eV  and   $ \Delta \mu_{\rm H_{2}O} < -0.91$ eV.
A more detailed description of the derivation of these quantities can be found elsewhere.\cite{fronzi2009-bis}

Equation \ref{gamma} describes the surface free energy, $\gamma$ for each of the surface-molecule systems, i.e. a function of the two variables $\Delta \mu_{\rm O_{2}}$ and $\Delta \mu_{\rm H_{2}O}$. For each value of $\Delta \mu_{\rm O_{2}}$ and $\Delta \mu_{\rm H_{2}O}$, the value of $\gamma$ with lower energy is the most thermodynamically stable.  Figure \ref{fig:stability} is obtained by projecting the surface with the lowest free energy on the $\Delta \mu_{\rm O_{2}}$$-$$\Delta \mu_{\rm H_{2}O}$ planes so that each colored area represents the most stable surface-molecule system for given values of the chemical potential of oxygen and water. 
In Fig. \ref{fig:stability}, the plot can be divided in two main areas: one including the blue and the red ones where CH$_4$ is stable on the reduced and stoichiometric surfaces, respectively, and the rest of them where the CH$_4$ is not stable. In these other areas CH$_x$ fragments are stable, indicating the environmental conditions where CH$_4$ will dissociate on the CeO$_2$(111) surface. 
As a general trend, for each fixed value of $\Delta \mu_{\rm O_{2}}$ the partial pressure of the water plays an important role in the dehydrogenation process.
In this case, the loss of one or more hydrogen atoms, and therefore the loss of one ore more electrons by the methane molecule, is equivalent to an oxidation process.  

Having defined the relative stability of each system, we now can consider under which conditions relevant transition phases occur.
Fixing the chemical potential of water imposing a water-rich condition ($\Delta \mu_{\rm H_{2}O}$=$-0.91$), for an atmospheric pressure $p$=1 atm, we find the surface to be reduced when the temperature is above $T$$\sim$1200 K, while for $p$=10$^{-14}$,  this value drops to $T$$\sim$500 K, indicating that the pressure of the oxygen play a fundamental role in changing the critical temperature for surface reduction.
Under atmospheric conditions, we can consider the partial pressure of the water vapor and oxygen to be around 1 atm. In this case the transition temperature between CeO$_2$+CH$_4$ and CeO$_2$+CH$_3$ is $\sim$750 K, while between CeO$_2$+CH$_3$ and CeO$_2$+CH is $T$$\sim$1050 K. This result suggests that a temperature above 750 K in the atmosphere is a sufficient condition for the partial oxidation of methane on the CeO$_2$(111) surface which is in a good agreement with Haneda \textit {et al.} who confirmed CeO$_2$ to show a catalytic activity for methane oxidation at 673 K.\cite{haneda1998} 
The difference could be due to the presence, supposedly small, of other facets rather then the more stable (111) in the specimen. 
As shown in Fig.\ref{fig:stability}, there is a competing mechanism played by oxygen and water chemical potentials against methane dissociation. An increment of $\Delta \mu_{\rm H_{2}O}$ would result in a higher critical temperature for CH$_4$ oxidation, while a change of $\Delta \mu_{\rm O_{2}}$ would have the opposite effect. This result qualitatively confirms the effect of water on methane oxidation observed experimentally on several other materials.\cite{cullis1984, ribeiro1994}

\section{Electronic structure}

\begin{figure}[!tbp]
\scalebox{0.20}{\includegraphics{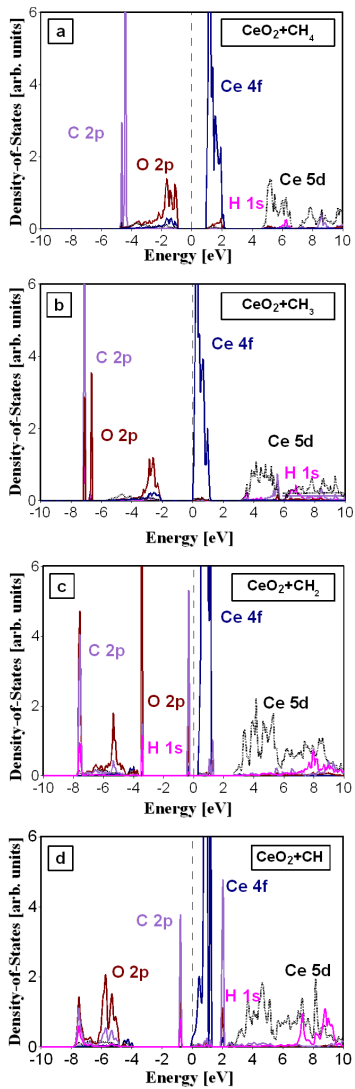}}
\caption{\label{fig:5} Projected density-of-states of CH$_x$ adsorbed on CeO$_2$(111) surface. (a), (b), (c) and (d) represents the CH$_4$, CH$_3$, CH$_2$ and CH on the CeO$_2$(111) surface. The black and the solid dark gray (blue) lines represent, respectively, the $5d$ and the $4f$ states of the cerium atoms, while the gray (red) line represents the $2p$ state of the oxygen atoms. The pale gray (magenta and purple) lines represent, respectively, the hydrogen $1s$ and carbon $2p$ of the CH$_x$ fragment. The vertical dashed line represents the Fermi energy.}
\end{figure}
 

In this section, we analyze the electronic structures of the most stable surfaces identified in the previous section (Figs.\,\ref{fig:1}, \,\ref{fig:2}, \,\ref{fig:3} and \,\ref{fig:4}).

Figure\,\ref{fig:5}(a) shows the projected density-of-states (pDOS) for a stoichiometric surface with adsorbed methane molecule. There is no evidence of hybridization in the states of the carbon atom of the methane molecule and the nearest surface atoms. 
When focusing on the surface atoms, we observe that the CeO$_2$(111) pDOS exhibits a strong localization of the Ce-$4f$ band, typical of strongly correlated materials, centered at 1 eV, while the Ce-$5d$ states are delocalized in the range of $5$ to $10$ eV. The highest occupied band is mainly derived from the $2p$ state of oxygen, which has an overlap in energy with the Ce-$4f$ and Ce-$5d$ states in the range $-5$ eV to $-1$ eV.  
The $\rm C_{\rm CH_4}$-$2p$ peaks preserve their narrow profile, indicating a molecular-like behavior.

For the case of CH$_3$ chemisorbed on CeO$_2$(111) (see Fig.\,\ref{fig:5}(b)), the ${\rm C}_{\rm CH_3}$-$2p$ peaks shift toward lower energies by 2 eV. 
The ${\rm O}_{\rm CeO_2}$-$2p$ shows a splitting in two narrow peaks and there is an evident overlapping between ${\rm C}_{\rm CH_3}$-$2p$ and ${\rm O}_{\rm CeO_2}$-$2p$ at around $-$7 eV, indicating an evident surface-molecule interaction.
This behavior  is typical of atomic-like orbitals interacting with a narrow band, in this specific case the peak of ${\rm O}_{\rm CeO_2}$-$2p$.\cite{norskov1990} 
In this case, the CeO$_2$(111) surface shows a metallic behavior. However, the occupation of the $4f$ band is very
low, and the energy of the surface can be considered indistinguishable to the semi-conductive surface. 

When CH$_2$ adsorbe on the surface (see Fig.\,\ref{fig:5}(c)), the hybridization of the orbitals is more evident.  The ${\rm O}_{\rm CeO_2}$-$2p$ band shows a further shift to $-$6 eV and a broadening one of the peak. There is also an evident overlapping of ${\rm O}_{\rm CeO_2}$-$2p$, ${\rm O}_{\rm CH_2}$-$2p$ and ${\rm H}_{\rm CH_2}$-$2p$ at around $-$8 eV and $-$4 eV, indicating a strong interaction.

When the fragment CH is adsorbed on the surface (see Fig.\,\ref{fig:5}(d)), the narrow peak of the ${\rm C}_{\rm CeO_2}$-$2p$ exhibits a broadening, typical of atomic-like orbitals interacting with a broad band.\cite{norskov1990} This feature
and suggests chemisorption of the adsorbate on the surface, that interacts with the band structure at surface level, consistent with the adsorption energy  value previously shown.  
Also, in the adsorption of the CH fragment, the surface shows a metallic behavior. However, analogous considerations on the energy of the system can be done as in the case of the CH$_3$ adsorption. 

\section{Summary and Conclusions}

We performed DFT calculations in order to have an insight to the
structure and energetics of CH$_x$ adsorption on the stoichiometric
and reduced CeO$_2$(111) surface for a coverage of
0.25 ML. Our results show that, on the stoichiometric
surface, the most stable structure is the one in which methane is
adsorbed on Ce. However, the adsorption energy indicate a very weak interaction. Also, we found that surface oxygen vacancies
do not increase the strength of the interaction, and that adsorption energies become stronger when methane is dehydrogenated.   
From the surface phase diagram, constructed by employing the \emph{ab initio} constrained thermodynamics approach, we studied the relevant phase transitions.
Fixing the chemical potential of water and imposing a water-rich condition, when the pressure of the atmosphere is $p$=1 atm, the reduced CeO$_2$(111) surface is thermodynamically stable for value of temperature is above $T$$\sim$1200 K, while for $p$=10$^{-14}$ this value drops to $T$$\sim$500 K, indicating the fundamental role of the oxygen in determining the critical temperature for the reduction of the surface.
Also, we estimated the sufficient conditions to obtain partial oxidation of the methane to be, considering the partial pressure of both water vapor and oxygen gas at 1 atm, 750 K, in good agreement with experimental results.  
A competing mechanism played by oxygen and water chemical potentials against methane dissociation was found, and while an increment of $\Delta \mu_{\rm H_{2}O}$ would result in a higher critical temperature for CH$_4$ oxidation, a change of $\Delta \mu_{\rm O_{2}}$ would have the opposite effect. This result qualitatively confirms the effect of water on methane oxidation observed experimentally on several other material surfaces.

\begin{acknowledgments}

The authors gratefully acknowledge support from the Australian Research Council (ARC), the Australian National Computational Infrastructure (NCI) and the Australian Center for Advanced Computing and Communication (AC3). MF expresses his gratitude to Jason Gao and Irina Holca for proofreading the manuscript.

\end{acknowledgments}



\bibliography{ceo2-met}

\begin{thebibliography}{30}
\expandafter\ifx\csname natexlab\endcsname\relax\def\natexlab#1{#1}\fi
\expandafter\ifx\csname bibnamefont\endcsname\relax
  \def\bibnamefont#1{#1}\fi
\expandafter\ifx\csname bibfnamefont\endcsname\relax
  \def\bibfnamefont#1{#1}\fi
\expandafter\ifx\csname citenamefont\endcsname\relax
  \def\citenamefont#1{#1}\fi
\expandafter\ifx\csname url\endcsname\relax
  \def\url#1{\texttt{#1}}\fi
\expandafter\ifx\csname urlprefix\endcsname\relax\def\urlprefix{URL }\fi
\providecommand{\bibinfo}[2]{#2}
\providecommand{\eprint}[2][]{\url{#2}}

\bibitem[{\citenamefont{Long et~al.}(2013)\citenamefont{Long, Yang, Thi, and
  N.V. Minhand Y. Caoand M.~Nogami}}]{long2013}
\bibinfo{author}{\bibfnamefont{N.}~\bibnamefont{Long}},
  \bibinfo{author}{\bibfnamefont{Y.}~\bibnamefont{Yang}},
  \bibinfo{author}{\bibfnamefont{C.~M.} \bibnamefont{Thi}}, \bibnamefont{and}
  \bibinfo{author}{\bibfnamefont{M.}~\bibnamefont{N.V. Minhand Y. Caoand
  M.~Nogami}}, \bibinfo{journal}{Nano Energy} \textbf{\bibinfo{volume}{2:5}},
  \bibinfo{pages}{636} (\bibinfo{year}{2013}).

\bibitem[{\citenamefont{Kendrick and Slater}(2013)}]{kendrick2013}
\bibinfo{author}{\bibfnamefont{E.}~\bibnamefont{Kendrick}} \bibnamefont{and}
  \bibinfo{author}{\bibfnamefont{P.~R.} \bibnamefont{Slater}},
  \bibinfo{journal}{Ann. Rep. Prog. Chem. A} \textbf{\bibinfo{volume}{109}},
  \bibinfo{pages}{139} (\bibinfo{year}{2013}).

\bibitem[{\citenamefont{Moussounda et~al.}(2007)\citenamefont{Moussounda,
  Haroun, Rakotovelo, and L\'{e}gar\'{e}}}]{Moussounda2007}
\bibinfo{author}{\bibfnamefont{P.}~\bibnamefont{Moussounda}},
  \bibinfo{author}{\bibfnamefont{M.}~\bibnamefont{Haroun}},
  \bibinfo{author}{\bibfnamefont{G.}~\bibnamefont{Rakotovelo}},
  \bibnamefont{and}
  \bibinfo{author}{\bibfnamefont{P.}~\bibnamefont{L\'{e}gar\'{e}}},
  \bibinfo{journal}{Surf. Sci.} \textbf{\bibinfo{volume}{3697-3701}},
  \bibinfo{pages}{601} (\bibinfo{year}{2007}).

\bibitem[{\citenamefont{Monnerat et~al.}(2001)\citenamefont{Monnerat,
  Kiwi-Minsker, and Renken}}]{monnerat2001}
\bibinfo{author}{\bibfnamefont{B.}~\bibnamefont{Monnerat}},
  \bibinfo{author}{\bibfnamefont{L.}~\bibnamefont{Kiwi-Minsker}},
  \bibnamefont{and} \bibinfo{author}{\bibfnamefont{A.}~\bibnamefont{Renken}},
  \bibinfo{journal}{Chem. Eng. Sci.} \textbf{\bibinfo{volume}{56}},
  \bibinfo{pages}{633} (\bibinfo{year}{2001}).

\bibitem[{\citenamefont{Saraf et~al.}(2005)\citenamefont{Saraf, Wang,
  Shutthanandan, Zhang, Marina, Baer, Thevuthasan, Nachimuthu, and
  Lindle}}]{saraf2005}
\bibinfo{author}{\bibfnamefont{L.}~\bibnamefont{Saraf}},
  \bibinfo{author}{\bibfnamefont{C.~M.} \bibnamefont{Wang}},
  \bibinfo{author}{\bibfnamefont{V.}~\bibnamefont{Shutthanandan}},
  \bibinfo{author}{\bibfnamefont{Y.}~\bibnamefont{Zhang}},
  \bibinfo{author}{\bibfnamefont{O.}~\bibnamefont{Marina}},
  \bibinfo{author}{\bibfnamefont{D.}~\bibnamefont{Baer}},
  \bibinfo{author}{\bibfnamefont{S.}~\bibnamefont{Thevuthasan}},
  \bibinfo{author}{\bibfnamefont{P.}~\bibnamefont{Nachimuthu}},
  \bibnamefont{and} \bibinfo{author}{\bibfnamefont{D.}~\bibnamefont{Lindle}},
  \bibinfo{journal}{J. Mat. Res.} \textbf{\bibinfo{volume}{20}},
  \bibinfo{pages}{1295} (\bibinfo{year}{2005}).

\bibitem[{\citenamefont{Kim and Maier}(2004)}]{kim2004}
\bibinfo{author}{\bibfnamefont{S.}~\bibnamefont{Kim}} \bibnamefont{and}
  \bibinfo{author}{\bibfnamefont{J.}~\bibnamefont{Maier}}, \bibinfo{journal}{J.
  Eur. Ceram. Soc.} \textbf{\bibinfo{volume}{24}}, \bibinfo{pages}{1919}
  (\bibinfo{year}{2004}).

\bibitem[{\citenamefont{Trovarelli}(1996)}]{trovarelli1996}
\bibinfo{author}{\bibfnamefont{A.}~\bibnamefont{Trovarelli}},
  \bibinfo{journal}{Catal. Rev. Sci.} \textbf{\bibinfo{volume}{38}},
  \bibinfo{pages}{439} (\bibinfo{year}{1996}).

\bibitem[{\citenamefont{Trovarelli}(2002)}]{trovarelli2002}
\bibinfo{author}{\bibfnamefont{A.}~\bibnamefont{Trovarelli}},
  \bibinfo{journal}{Catal. Sci. Series} \textbf{\bibinfo{volume}{2}},
  \bibinfo{pages}{407} (\bibinfo{year}{2002}).

\bibitem[{\citenamefont{Farrauto}(2012)}]{farrauto2012}
\bibinfo{author}{\bibfnamefont{R.~J.} \bibnamefont{Farrauto}},
  \bibinfo{journal}{Science} \textbf{\bibinfo{volume}{337}},
  \bibinfo{pages}{659} (\bibinfo{year}{2012}).

\bibitem[{\citenamefont{Haneda et~al.}(1998)\citenamefont{Haneda, Mizushima,
  and Kakuta}}]{haneda1998}
\bibinfo{author}{\bibfnamefont{M.}~\bibnamefont{Haneda}},
  \bibinfo{author}{\bibfnamefont{T.}~\bibnamefont{Mizushima}},
  \bibnamefont{and} \bibinfo{author}{\bibfnamefont{N.}~\bibnamefont{Kakuta}},
  \bibinfo{journal}{J. Phys. Chem. B} \textbf{\bibinfo{volume}{102}},
  \bibinfo{pages}{6579} (\bibinfo{year}{1998}).

\bibitem[{\citenamefont{Laosiripojana et~al.}(2004)\citenamefont{Laosiripojana,
  Assabumrungrat, and Charojrochkul}}]{laosiripojana2004}
\bibinfo{author}{\bibfnamefont{N.}~\bibnamefont{Laosiripojana}},
  \bibinfo{author}{\bibfnamefont{S.}~\bibnamefont{Assabumrungrat}},
  \bibnamefont{and}
  \bibinfo{author}{\bibfnamefont{S.}~\bibnamefont{Charojrochkul}},
  \bibinfo{journal}{The Joint International Conference on Sustainable Energy
  and Environment}  (\bibinfo{year}{2004}).

\bibitem[{\citenamefont{Knapp and Ziegler}(2008)}]{knapp2008}
\bibinfo{author}{\bibfnamefont{D.}~\bibnamefont{Knapp}} \bibnamefont{and}
  \bibinfo{author}{\bibfnamefont{T.}~\bibnamefont{Ziegler}},
  \bibinfo{journal}{J. Phys. Chem. C} \textbf{\bibinfo{volume}{112}},
  \bibinfo{pages}{17311} (\bibinfo{year}{2008}).

\bibitem[{\citenamefont{McIntosh and Gorte}(2004)}]{mcintosh2004}
\bibinfo{author}{\bibfnamefont{S.}~\bibnamefont{McIntosh}} \bibnamefont{and}
  \bibinfo{author}{\bibfnamefont{R.~J.} \bibnamefont{Gorte}},
  \bibinfo{journal}{Chem. Rev.} \textbf{\bibinfo{volume}{104}},
  \bibinfo{pages}{4845} (\bibinfo{year}{2004}).

\bibitem[{\citenamefont{Murray et~al.}(1999)\citenamefont{Murray, Tsai, and
  Barnett}}]{murray1999}
\bibinfo{author}{\bibfnamefont{E.~P.} \bibnamefont{Murray}},
  \bibinfo{author}{\bibfnamefont{T.}~\bibnamefont{Tsai}}, \bibnamefont{and}
  \bibinfo{author}{\bibfnamefont{S.~A.} \bibnamefont{Barnett}},
  \bibinfo{journal}{Nature} \textbf{\bibinfo{volume}{400}},
  \bibinfo{pages}{649} (\bibinfo{year}{1999}).

\bibitem[{\citenamefont{Odier et~al.}(2007)\citenamefont{Odier, Schuurman, and
  Mirodatos}}]{odier2007}
\bibinfo{author}{\bibfnamefont{E.}~\bibnamefont{Odier}},
  \bibinfo{author}{\bibfnamefont{Y.}~\bibnamefont{Schuurman}},
  \bibnamefont{and}
  \bibinfo{author}{\bibfnamefont{C.}~\bibnamefont{Mirodatos}},
  \bibinfo{journal}{Cat. Today} \textbf{\bibinfo{volume}{127}},
  \bibinfo{pages}{230} (\bibinfo{year}{2007}).

\bibitem[{\citenamefont{Bharadwaj and Schmidt}(1995)}]{bharadwaj1995}
\bibinfo{author}{\bibfnamefont{S.}~\bibnamefont{Bharadwaj}} \bibnamefont{and}
  \bibinfo{author}{\bibfnamefont{L.}~\bibnamefont{Schmidt}},
  \bibinfo{journal}{Fuel Proc. Tech.} \textbf{\bibinfo{volume}{42}},
  \bibinfo{pages}{109} (\bibinfo{year}{1995}).

\bibitem[{\citenamefont{Cullis and Willatt}(1984)}]{cullis1984}
\bibinfo{author}{\bibfnamefont{C.~F.} \bibnamefont{Cullis}} \bibnamefont{and}
  \bibinfo{author}{\bibfnamefont{B.~M.} \bibnamefont{Willatt}},
  \bibinfo{journal}{J. Catal.} \textbf{\bibinfo{volume}{86}},
  \bibinfo{pages}{187} (\bibinfo{year}{1984}).

\bibitem[{\citenamefont{Ribeiro et~al.}(1994)\citenamefont{Ribeiro, Chow, and
  Dallabetta}}]{ribeiro1994}
\bibinfo{author}{\bibfnamefont{F.~H.} \bibnamefont{Ribeiro}},
  \bibinfo{author}{\bibfnamefont{M.}~\bibnamefont{Chow}}, \bibnamefont{and}
  \bibinfo{author}{\bibfnamefont{R.~A.} \bibnamefont{Dallabetta}},
  \bibinfo{journal}{J. Catal.} \textbf{\bibinfo{volume}{146}},
  \bibinfo{pages}{537} (\bibinfo{year}{1994}).

\bibitem[{\citenamefont{Kikuchi et~al.}(2002)\citenamefont{Kikuchi, Maeda,
  Sasaki, Wennerstr\"{o}mc, and Eguchi}}]{kikuchi2002}
\bibinfo{author}{\bibfnamefont{R.}~\bibnamefont{Kikuchi}},
  \bibinfo{author}{\bibfnamefont{S.}~\bibnamefont{Maeda}},
  \bibinfo{author}{\bibfnamefont{K.}~\bibnamefont{Sasaki}},
  \bibinfo{author}{\bibfnamefont{S.}~\bibnamefont{Wennerstr\"{o}mc}},
  \bibnamefont{and} \bibinfo{author}{\bibfnamefont{K.}~\bibnamefont{Eguchi}},
  \bibinfo{journal}{Appl. Cat. A} \textbf{\bibinfo{volume}{232}},
  \bibinfo{pages}{23} (\bibinfo{year}{2002}).

\bibitem[{\citenamefont{Mayernick and Janik}(2008)}]{mayernick2008}
\bibinfo{author}{\bibfnamefont{A.~D.} \bibnamefont{Mayernick}}
  \bibnamefont{and} \bibinfo{author}{\bibfnamefont{M.~J.} \bibnamefont{Janik}},
  \bibinfo{journal}{J. Phys. Chem. C} \textbf{\bibinfo{volume}{112}},
  \bibinfo{pages}{14955} (\bibinfo{year}{2008}).

\bibitem[{\citenamefont{Stampfl}(2005)}]{stampfl2005}
\bibinfo{author}{\bibfnamefont{C.}~\bibnamefont{Stampfl}},
  \bibinfo{journal}{Catal. Today} \textbf{\bibinfo{volume}{105}},
  \bibinfo{pages}{17} (\bibinfo{year}{2005}).

\bibitem[{\citenamefont{Reuter and Scheffler}(2002)}]{reuter2002}
\bibinfo{author}{\bibfnamefont{K.}~\bibnamefont{Reuter}} \bibnamefont{and}
  \bibinfo{author}{\bibfnamefont{M.}~\bibnamefont{Scheffler}},
  \bibinfo{journal}{Phys. Rev. B} \textbf{\bibinfo{volume}{65}},
  \bibinfo{pages}{035406} (\bibinfo{year}{2002}).

\bibitem[{\citenamefont{Reuter and Scheffler}(2003)}]{reuter2003}
\bibinfo{author}{\bibfnamefont{K.}~\bibnamefont{Reuter}} \bibnamefont{and}
  \bibinfo{author}{\bibfnamefont{M.}~\bibnamefont{Scheffler}},
  \bibinfo{journal}{Phys. Rev. B} \textbf{\bibinfo{volume}{68}},
  \bibinfo{pages}{045407} (\bibinfo{year}{2003}).

\bibitem[{\citenamefont{Stampfl et~al.}(2002)\citenamefont{Stampfl,
  Ganduglia-Pirovano, Reuter, and Scheffler}}]{stampfl2002}
\bibinfo{author}{\bibfnamefont{C.}~\bibnamefont{Stampfl}},
  \bibinfo{author}{\bibfnamefont{M.~V.} \bibnamefont{Ganduglia-Pirovano}},
  \bibinfo{author}{\bibfnamefont{K.}~\bibnamefont{Reuter}}, \bibnamefont{and}
  \bibinfo{author}{\bibfnamefont{M.}~\bibnamefont{Scheffler}},
  \bibinfo{journal}{Surf. Sci.} \textbf{\bibinfo{volume}{500}},
  \bibinfo{pages}{368} (\bibinfo{year}{2002}).

\bibitem[{\citenamefont{Perdew et~al.}(1996)\citenamefont{Perdew, Burke, and
  Erzerhof}}]{perdew1996}
\bibinfo{author}{\bibfnamefont{J.~P.} \bibnamefont{Perdew}},
  \bibinfo{author}{\bibfnamefont{K.}~\bibnamefont{Burke}}, \bibnamefont{and}
  \bibinfo{author}{\bibfnamefont{M.}~\bibnamefont{Erzerhof}},
  \bibinfo{journal}{Phys. Rev. Lett} \textbf{\bibinfo{volume}{77}},
  \bibinfo{pages}{3865} (\bibinfo{year}{1996}).

\bibitem[{\citenamefont{Delley}(1990)}]{delley1990}
\bibinfo{author}{\bibfnamefont{B.}~\bibnamefont{Delley}},
  \bibinfo{journal}{\jcp} \textbf{\bibinfo{volume}{92}}, \bibinfo{pages}{508}
  (\bibinfo{year}{1990}).

\bibitem[{\citenamefont{Delley}(2000)}]{delley2000}
\bibinfo{author}{\bibfnamefont{B.}~\bibnamefont{Delley}},
  \bibinfo{journal}{\jcp} \textbf{\bibinfo{volume}{113}}, \bibinfo{pages}{7756}
  (\bibinfo{year}{2000}).

\bibitem[{\citenamefont{Fronzi et~al.}(2009{\natexlab{a}})\citenamefont{Fronzi,
  Soon, Delley, Traversa, and Stampfl}}]{fronzi2009}
\bibinfo{author}{\bibfnamefont{M.}~\bibnamefont{Fronzi}},
  \bibinfo{author}{\bibfnamefont{A.}~\bibnamefont{Soon}},
  \bibinfo{author}{\bibfnamefont{B.}~\bibnamefont{Delley}},
  \bibinfo{author}{\bibfnamefont{E.}~\bibnamefont{Traversa}}, \bibnamefont{and}
  \bibinfo{author}{\bibfnamefont{C.}~\bibnamefont{Stampfl}},
  \bibinfo{journal}{J. Chem. Phys.} \textbf{\bibinfo{volume}{131}},
  \bibinfo{pages}{104701} (\bibinfo{year}{2009}{\natexlab{a}}).

\bibitem[{\citenamefont{Fronzi et~al.}(2009{\natexlab{b}})\citenamefont{Fronzi,
  Piccinin, Delley, Traversa, and Stampfl}}]{fronzi2009-bis}
\bibinfo{author}{\bibfnamefont{M.}~\bibnamefont{Fronzi}},
  \bibinfo{author}{\bibfnamefont{S.}~\bibnamefont{Piccinin}},
  \bibinfo{author}{\bibfnamefont{B.}~\bibnamefont{Delley}},
  \bibinfo{author}{\bibfnamefont{E.}~\bibnamefont{Traversa}}, \bibnamefont{and}
  \bibinfo{author}{\bibfnamefont{C.}~\bibnamefont{Stampfl}},
  \bibinfo{journal}{Phys. Chem. Chem. Phys.} \textbf{\bibinfo{volume}{11}},
  \bibinfo{pages}{9188–9199} (\bibinfo{year}{2009}{\natexlab{b}}).

\bibitem[{\citenamefont{N\o{}rskov}(1990)}]{norskov1990}
\bibinfo{author}{\bibfnamefont{J.~K.} \bibnamefont{N\o{}rskov}},
  \bibinfo{journal}{Rep. Prog. Phys.} \textbf{\bibinfo{volume}{53}},
  \bibinfo{pages}{1253} (\bibinfo{year}{1990}).

\end{thebibliography}

\end{document}